\documentclass[conference]{IEEEtran}
\IEEEoverridecommandlockouts

\usepackage{amsmath,amssymb,amsfonts}
\usepackage{algorithmic}
\usepackage{graphicx}
\usepackage{textcomp}
\usepackage{xcolor}
\usepackage[english]{babel}
\usepackage{amsthm}

\usepackage{biblatex}
\theoremstyle{definition}
\newtheorem{definition}{Definition}[section]
    
\begin{document}

\title{Shadow Blade: A tool to interact with attack vectors}

\author{\IEEEauthorblockN{Ariel R. Ril\textsuperscript{1}, Daniel Dalalana Bertoglio\textsuperscript{2} and Avelino F. Zorzo\textsuperscript{3}}
\IEEEauthorblockA{\textit{Pontifical Catholic University of Rio Grande do Sul}\\
\textit{School of Technology}\\
Porto Alegre, Brazil \\
ariel.ril@edu.pucrs.br\textsuperscript{1}, daniel.bertoglio@edu.pucrs.br\textsuperscript{2} and avelino.zorzo@pucrs.br\textsuperscript{3}}
}

\maketitle

\begin{abstract}
 The increased demand of cyber security professionals has also increased the development of new platforms and tools that help those professionals to improve their offensive skills. One of these platforms is HackTheBox, an online cyber security training platform that delivers a controlled and safe environment for those professionals to explore virtual machines in a Capture the Flag (CTF) competition style. 
 
 Most of the tools used in a CTF, or even on real-world Penetration Testing (Pentest), were developed for specific reasons so each tool usually has different input and output formats. These different formats make it hard for cyber security professionals and CTF competitors to develop an attack graph. In order to help cyber security professionals and CTF competitors to discover, select and exploit an attack vector, this paper presents Shadow Blade, a tool to aid users to interact with their attack vectors.
 
\end{abstract}


\section{Introduction}\label{sec:intro}

Computer technology has been spread throughout all areas in modern society. This wide spread use has brought many advantages to people, helping to improve the productivity of their activities. Despite all these advantages, this new environment has also increased the number of virtual attacks that are being deployed in the Digital World. 

In order to reduce the impact of these attacks, new skills to Information Technology (IT) professionals are needed. These skills are related to understand the types of attacks that may affect people in real life. Two main types of skill can be described as defensive and offensive. In the former, a ``reactive approach to security that focuses on prevention, detection, and response to attacks" is used. In the latter, ``a proactive approach to security through the use of ethical hacking'' is applied \cite{understand-cybersec-tracks}. This paper is related to the offensive strategy.

In the cyber security universe there are many ways to sharpen the offensive skills \cite{comptia-learn} and one of these ways is a Capture The Flag (CTF) Competition \cite{ctf-ascyberintro}. The CTF Competition objective is to find flags (pieces of data) that are inside specially crafted virtual machines (VM). These VMs are developed with selected vulnerabilities so the competitors can explore and exploit these vulnerabilities to access the VM and \textit{steal} those flags. 


In the processes of exploring a machine in a CTF, the competitor must use a comprehensive list of tools for reconnaissance (\textit{e.g.}, nmap \cite{nmap}, ffuf \cite{ffuf}), vulnerability assessment (searchsploit \cite{searchsploit}) and exploitation (metasploit \cite{metasploit}). But for the competitors to reach the exploitation step, they have to create an artifact from the previous steps to be able to create an attack graph, which will be used to filter and select the best attack vector to be followed in order to compromise the machine. The attack graph is the data structure used to represent all possible attacks on a computer network, in which one or more attack vectors will be selected to successfully exploit a target.

Most of the tools that are used during a CTF, or even in the real-world for a Penetration Testing (Pentest) \cite{Dalalana2017}, were created for a specific purpose. Hence, each tool usually has different input and output formats. These different formats of generated data make it hard for the CTF competitors and cyber security professionals to compile and analyze the data and develop the artifact that is needed to select the best attack vector. 

Therefore, in order to help cyber security professionals and CTF competitors to discover, select and exploit an attack vector, this paper presents Shadow Blade, a tool to aid users to interact with their attack vectors. Shadow Blade is a tool that can help users by running reconnaissance tools (nmap and ffuf, initially) and develop a visualization of attack vectors in the intended target using directed graphs. 

This paper is organized as follows. In Section \ref{sec:background} we present the background for this work. Section \ref{sec:related} presents the related work. Section \ref{sec:proposal} describes our proposal. Section \ref{sec:sim-tools} describes similar tools and compares them with Shadow Blade. Section \ref{sec:evaluation} presents some preliminary evaluation of our tool. And, finally, Section \ref{sec:conclusion} and Section \ref{sec:future} present our conclusions and future work.

\section{Background}
\label{sec:background}

In this section we present some topics that are related to our proposal. Initially, in Section \ref{subsec-attack-vec}, we define what are attack graphs and attack vectors. Afterwards, in Section \ref{subsec-ctf}, we describe how a CTF works. In Section \ref{subsec-htb} we describe how HackTheBox, the platform that we tested Shadow Blade, works. Finally, in Section \ref{subsec-digraph} we define the base structure that Shadow Blade uses. 

\begin{figure*}
    \centering
    \includegraphics[width=0.7\textwidth]{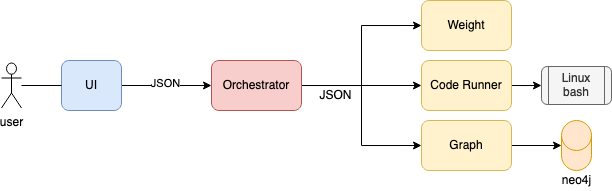}
    \caption{Shadow Blade architecture}
    \label{fig:shdw-arch}
\end{figure*}

\subsection{Attack Graph \& Attack Vectors} \label{subsec-attack-vec}

An attack graph is the data structure used to represent all possible attacks on a network \cite{1004377} and can be formally defined as:

\begin{definition}
An \textit{attack graph} (AG) is a tuple $G = (S, \tau, S_0, S_s)$, where $S$ is a set of states, $\tau \subseteq S \times S$ is a transition relation, $S_0 \subseteq S$ is a set of initial states, and $S_s \subseteq S$ is a set of success states.
\end{definition}

An attack vector is a sequence of actions or techniques that could compromise a machine, an enterprise or a computer network. Attack vectors are a subset of an attack graph.

\subsection{Capture the Flag (CTF)} \label{subsec-ctf}

A Capture the Flag (CTF) Competition is, usually, an online competition where the competitors have a list of vulnerable virtual machines, which are created to be intentionally exploited so the competitors can improve their offensive skills. The objective of a CTF is to compromise the bigger number of machines as possible, exploiting their vulnerabilities to retrieve flags (pieces of data or random strings) that are hidden in the machine.

\subsection{HackTheBox (HTB)} \label{subsec-htb}

HackTheBox \cite{htb:2021} is an online cyber security training platform that presents a long list of vulnerable machines, a list of security challenges and real-world scenarios so their users can always improve their security skills. Each vulnerable machine in the platform is categorized using a difficulty level that can be easy, medium, hard or insane. 

An easy machine at HTB commonly has a very explicit attack vector, so less experienced users can learn the basics of cyber security and the basics of offensive skills. Nonetheless, an insane machine is closest to a real-world scenario that could be an enterprise application that was almost cleared of vulnerabilities, but using a complex attack graph is still possible to compromise it.

\subsection{Directed Graph} \label{subsec-digraph}

A directed graph \cite{graph-theory} is a graph made of a set of vertices connected with directed edges. Formally defined as $G = (V, E)$ where:

\begin{itemize}
    \item $V$ is a set of vertices, nodes or points
    \item $E$ is a set of ordered pair of vertices, which define the start and end of an edge
\end{itemize}

\section{Related Work}
\label{sec:related}

Obes, Sarrute \& Richarte \cite{attk-plan-real-world} developed a tool that receive as input a computer network definition and executes a planning process using \textit{Planning Domain Definition Language} (PDDL) and uses SGPlan \cite{sgplan} and Metric-FF \cite{metric-ff} as planners to evaluate the PDDL plan. PDDL is used to define the actions that need to be executed (exploit execution, reconnaissance, ...) and to identify the final goal, which is to fully compromising the computer network achieving the highest access level. Therefore, it is possible to verify the possibility to automate all exploration process in a computer network. Nonetheless,  despite the fact that it was used in a realistic computer network, this approach has an exponential complexity growth directly proportional to the number of hosts in the network. The complexity growth is due to each new machine in the network have a connection to all previous machines, so the planner has to evaluate all the new paths for possible vulnerabilities.

Amman \& Pamula \cite{host-based-net-attk} approached this problem using a simplified definition for the planner execution and the automation of the exploration process. This simplification was in the execution of the automated exploration. The developed algorithm analyses only the connections from the current machine to decide which plan to execute, it is not necessary to analyse the complete network. Therefore, this simplified approach implies that the protective measures to mitigate the security issues against machine exploitation are sub-optimal.

\section{Proposal: Shadow Blade}
\label{sec:proposal}

In this paper we discuss Shadow Blade, a tool to interact with  attack vectors that is a modular tool developed to help cyber security professionals and CTF competitors on executing reconnaissance tools (\texttt{nmap} or \texttt{ffuf}, in the first version) and afterwards create a visualization with directed graphs of the attack vectors that were found. This section presents the architecture and development decisions of Shadow Blade.

\subsection{Architecture}

The Shadow Blade architecture follows the Domain-Driven Design (DDD) \cite{evans2004ddd}. In DDD each component is defined based in a domain, which could be defined as a context that the component will have knowledge of.

Each component, or module, of Shadow Blade executes only one functionality in order to allow the possibility of replacing a complete module without harming the whole system. The developed modules for this first version can be visualized in Figure \ref{fig:shdw-arch}:

\begin{itemize}
    \item \textbf{Weight}: An API that given a graph can compute the weight of each edge and returns the result;
    \item \textbf{Code Runner}: A façade \cite{gamma1994design} API for Linux Shell \cite{gnu2007free};
    \item \textbf{Graph}: An API that manages the data of an attack graph using Neo4j \cite{neo4j}, a graph database;
    \item \textbf{Orchestrator}: A façade for all raw functionality of Shadow Blade;
    \item \textbf{UI}: A React \cite{reactjs} app that the user can interact with and view results from \texttt{nmap} and \texttt{ffuf}
\end{itemize}

Different languages were used to develop each module. The language choice was based on each use case, so it could be flexible to expand and easy enough to understand the project organization.

\subsection{Weight}

The Weight API was developed using Go \cite{golang}, a language whose format is similar to a low level programming language, but it has high level programming languages constructs. \textit{Go is expressive, concise, clean, and efficient} \cite{godoc-frase}.

For the HTTP request handling, the Gin Web framework \cite{go-gin} was used. Gin is one of the most popular web frameworks for Go, it is fast, modular and it has the ability to add middleware in the process of handling a request to apply a functionality to all or some HTTP endpoints.

\subsection{Code Runner} \label{code-runner}

This module was developed using Go and Gin, but it has the ability to transform an HTTP request into commands for a Linux Shell. The first version of our tool uses  \texttt{nmap} \cite{nmap} and \texttt{ffuf} \cite{ffuf}. The reason for that is because \texttt{nmap} is one of the main tools for network scanning and \texttt{ffuf} is able to perform a series of web scanning varying from path discovery to API fuzzing.

\subsection{Graph}

The Graph module was developed using Typescript \cite{typescript}, a variation of JavaScript that is typed. The choice of Typescript was because the language forces the type definition and this definitions lower the possibility to have some type related bug in the program. ExpressJS \cite{expressjs} was chosen to be the web application framework because it is easy to setup, easy to understand and modular.

This is the main module of the project because it holds the information of the attack graph in a Neo4j \cite{neo4j} database: a database that is designed to save the data as nodes and edges. We chose Neo4j because the application result produce a directed graph in which nodes represent network ports or HTTP endpoints and edges contain the probability that a step from and to a node would have when an attack path could be exploited.

\subsection{Orchestrator}

This module coordinates the actions selected by the user in the user interface (UI). The Orchestrator was develop with NodeJS \cite{nodejs} using ExpressJS as the web framework and it is mainly divided into \textit{target} and \textit{graph}. The \textit{target} resource will be the machine that the user is exploring in a CTF and the \textit{graph} will be the attack graph generated from the output of execution of the tools implemented in the Code Runner (see Section \ref{code-runner}).

\subsection{User Interface (UI)}
This is the user interface where the user can register a new target in the platform and execute actions against the target to start the exploration process and receive a visual feedback that is the generated attack graph. The implementation of the module uses ReactJS \cite{reactjs}, which is a declarative library used to develop frontend applications using a component based approach.

\section{Similar Tools}
\label{sec:sim-tools}

The cyber security community has developed some tools that creates a visualization of a computer network using data generated from previous scans. Maltego \cite{maltego} that "is an open source intelligence and graphical analysis tool for gathering and connecting information for investigative tasks" generates a visualization of a computer network using information obtained from different reconnaissance sources, which can overwhelm the user with information. On the other hand Shadow Blade generates a simplified directed graph, so the user can learn from the information gathered in Shadow Blade and create its own directed graph. 

\begin{figure}
    \centering
    \includegraphics[width=8.5cm]{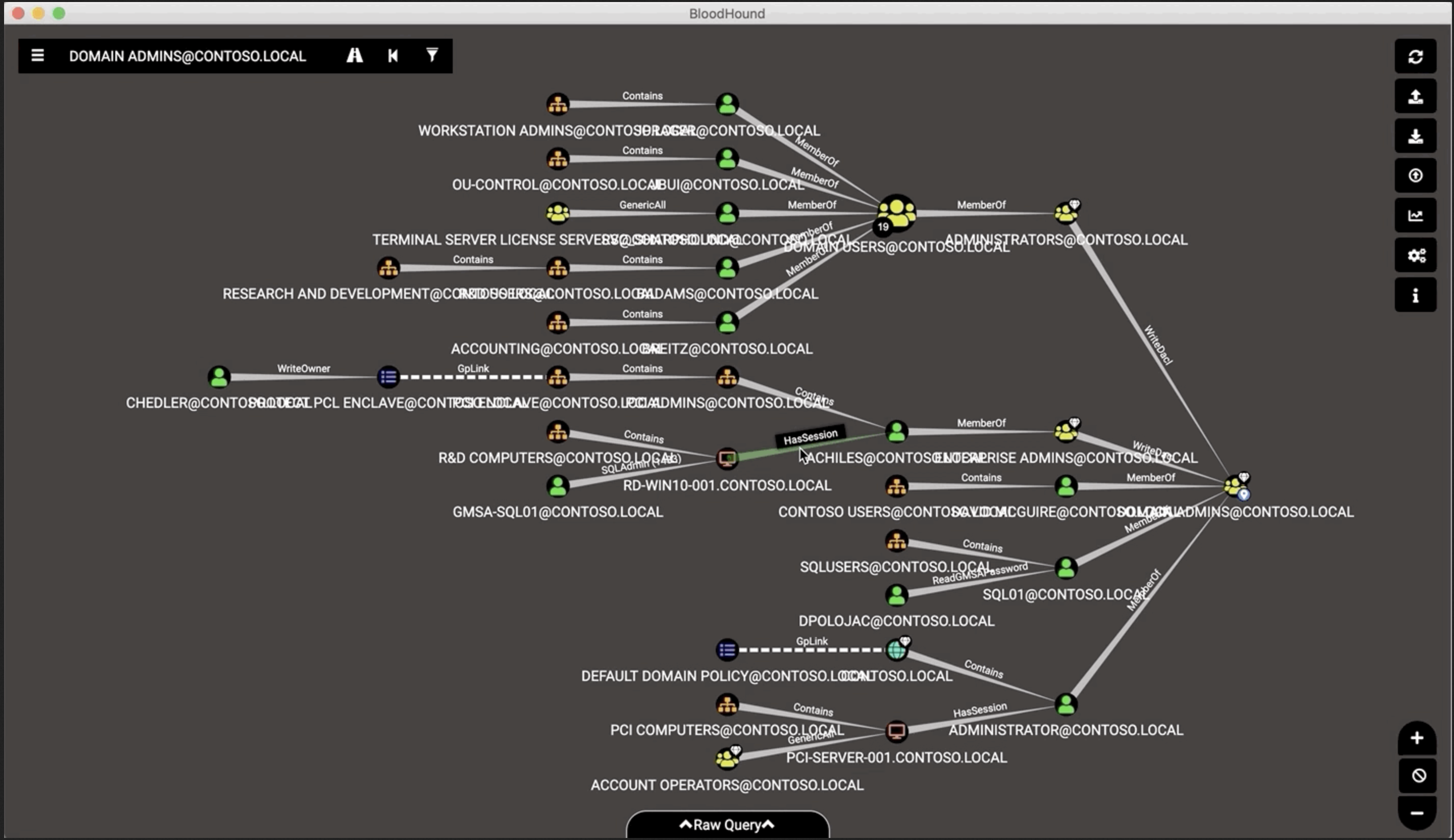}
    \caption{BloodHound showing the connections of objects in an Active Directory}
    \label{fig:bloodhound}
\end{figure}

BloodHound \cite{bloodhound} gathers information from an Active Directory (AD) \cite{active-directory} and generates a directed graph of the connections of each object in the AD. In Figure \ref{fig:bloodhound} we show BloodHound is used to extract information from an AD. The main difference from BloodHound to Shadow Blade is that Shadow Blade generates an attack graph with information of any computer network while BloodHound can only be used with an AD.

\section{Evaluation}
\label{sec:evaluation}

To evaluate if Shadow Blade would help cyber security professionals and CTF competitors to execute reconnaissance tools, gather the output data and create a visualization of attack vectors, we used an out-of-use machine in HackTheBox (HTB). The choice of that machine in HTB was that out-of-use machines have walk through on how to explore and exploit these machines and compare the results obtained in Shadow Blade with the actual attack vector of the machine.

The chosen machine for the evaluation was the out-of-use machine Armageddon, which is rated as easy in the platform. This machine has a vulnerable Drupal CMS \cite{drupal-cms} - a content management system - that is vulnerable to Remote Code Execution (RCE) \cite{drupalgeddon-cve}, popularly called \textit{Drupalgeddon}.



\begin{figure}
    \centering
    \includegraphics[width=8.5cm]{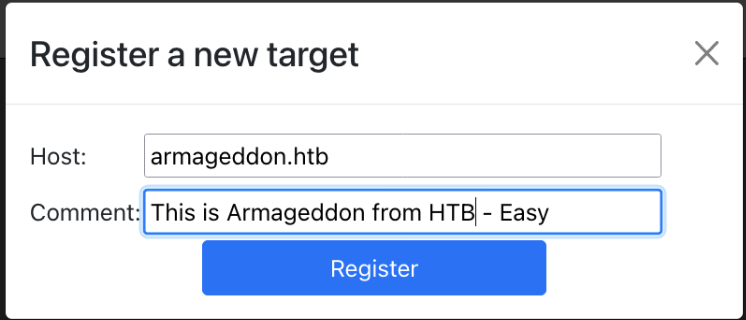}
    \caption{Registering Armageddon machine in Shadow Blade}
    \label{fig:target-register}
\end{figure}

After initializing all Shadow Blade modules the user can access the UI and start registering the intended target. To register a new target the user can use the target IP or a custom host pre-configured on the user machine. The process of registering a new target is showed in  Figure \ref{fig:target-register}, in the example we used a custom host for Armageddon (\textit{armageddon.htb}).

\begin{figure}
    \centering
    \includegraphics[width=8.5cm]{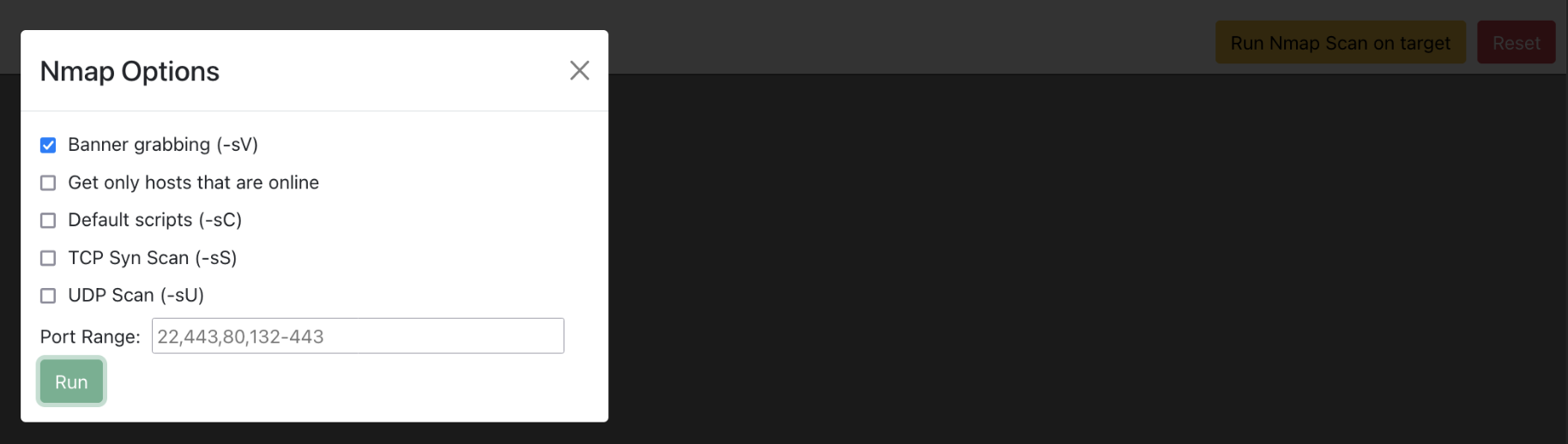}
    \caption{Selection of \texttt{nmap} options}
    \label{fig:armageddon-nmap-select}
\end{figure}

With the target registered, a user can start the reconnaissance processes executing \texttt{nmap} to discover the open ports of the machine and what services are running in each port. For this, the user can use the "Run Nmap scan on Target" button and select which \texttt{nmap} options to use (Figure \ref{fig:armageddon-nmap-select}). For this project we enabled some \texttt{nmap} options: 

\begin{itemize}
    \item \textbf{Banner grabbing}: is the process to connect to a certain port in a machine and retrieve the banner (data) that the service executing in the port returns to any attempt of connection;
    \item \textbf{Hosts online}: \texttt{nmap} has the ability to connect to a machine port and verify whether the port is online (returning information of a successful connection), or the port is filtered (the machine is behind a firewall and the firewall is dropping the attempts to connect in a port), or the port is closed;
    \item \textbf{Default scripts}: with the port number and the banner information, \texttt{nmap} can select default scripts that the community developed and execute against the machine. Usually these scripts are used for identification of vulnerable services and service default configuration.
\end{itemize}

\begin{figure}
    \centering
    \includegraphics[width=8.7cm]{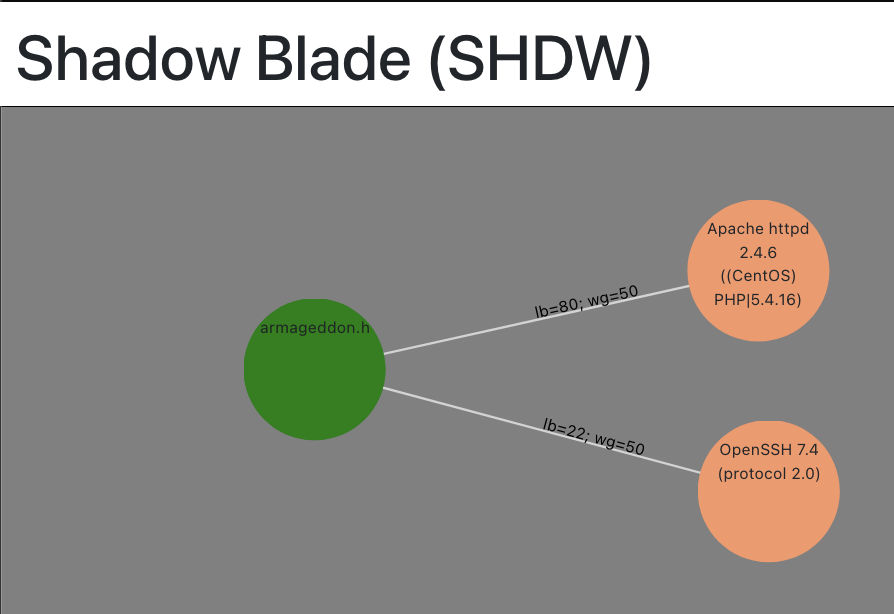}
    \caption{Result of executing nmap against Armageddon from HTB}
    \label{fig:armageddon-graph}
\end{figure}

As shown in Figure \ref{fig:armageddon-graph}, ports 22 and 80 in the Armageddon machine are  open: in port 80, an Apache server version 2.4.6 in a CentOS is running; and in port 22, OpenSSH 7.4 is running. With this information, the user can click in the Apache node and select some options to execute \texttt{ffuf} on port 80 of the machine.

\begin{figure}
    \centering
    \includegraphics[width=8.7cm]{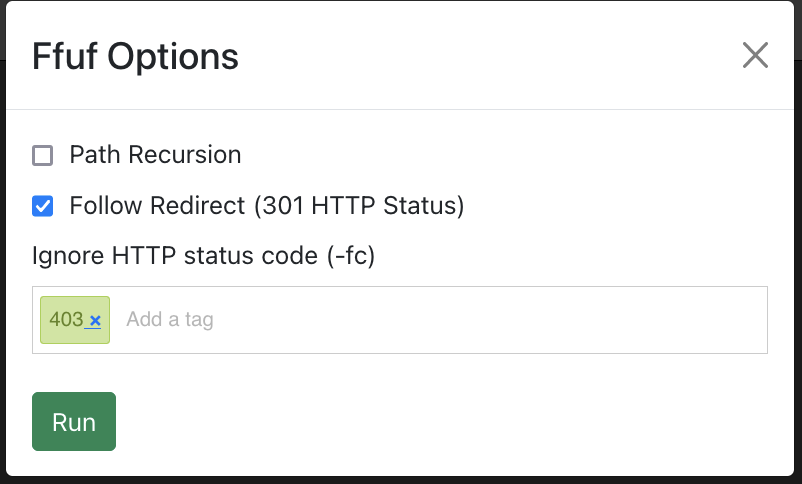}
    \caption{Options to execute \texttt{ffuf}}
    \label{fig:armageddon-ffuf-opts}
\end{figure}

Clicking in the Apache node will render a window with the options to execute \texttt{ffuf}, which are shown in Figure \ref{fig:armageddon-ffuf-opts}. In this example we configured two options to execute \texttt{ffuf}: to follow redirects and remove all  ``HTTP Status Code returns 403" (forbidden) results. The options developed for this project were:

\begin{itemize}
    \item \textbf{Path recursion}: \texttt{ffuf} will execute a recursion in the folders that it finds. For example, if a web server returns that the path "/js" exists than \texttt{ffuf} will attempt to find more directories or files in the path "/js/", recursively;
    \item \textbf{Follow redirect}: \texttt{ffuf} will follow the 301 HTTP Status Code, which defines a redirect from the current web page, sending the user the a new location;
    \item \textbf{Ignore HTTP Status Code}: \texttt{ffuf} will ignore every HTTP response that have the Status Code present in the configured list.
\end{itemize}

\begin{figure*}
    \centering
    \includegraphics[width=\linewidth]{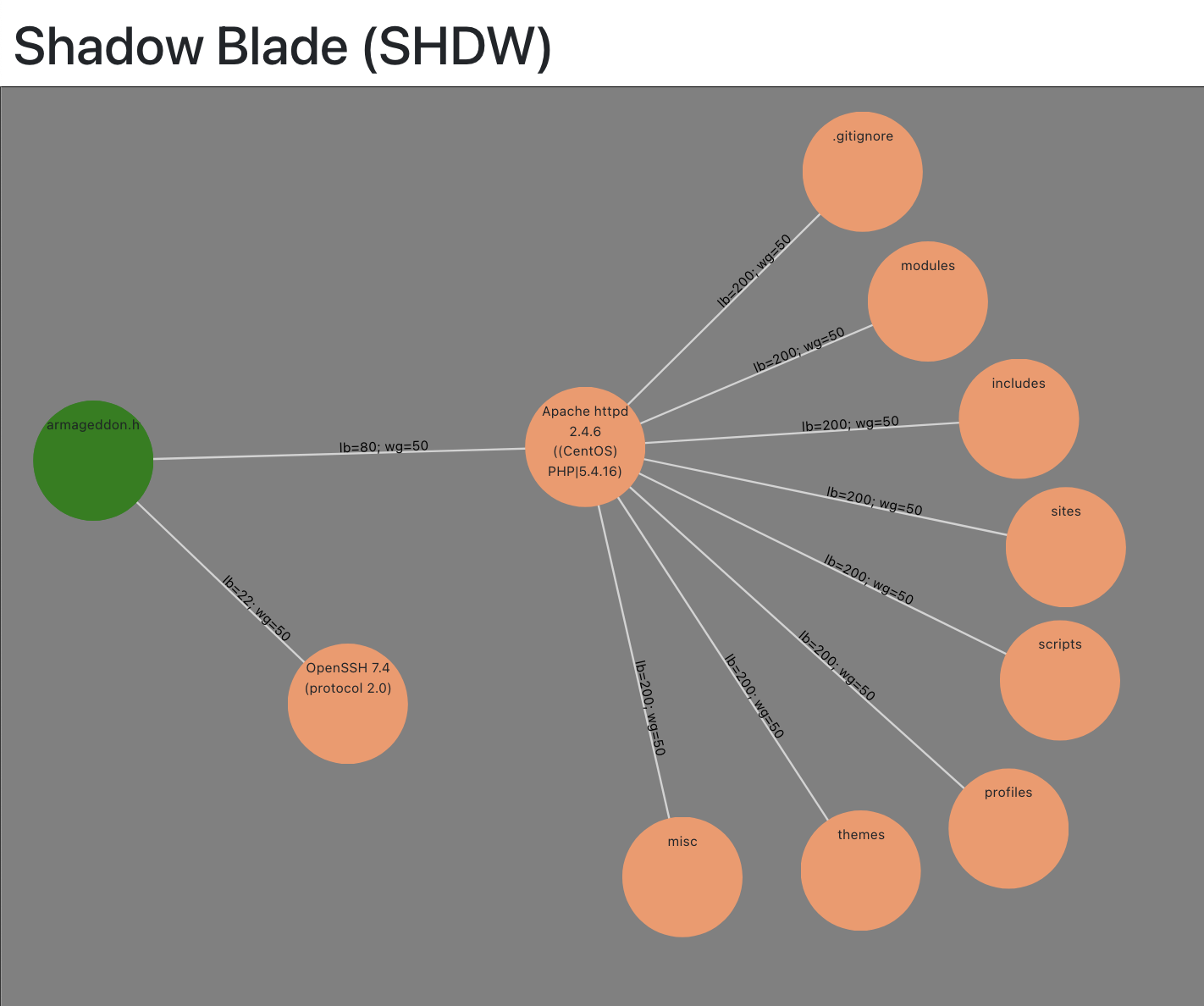}
    \caption{Result attack graph from executing \texttt{ffuf} in port 80 from Armageddon}
    \label{fig:armageddon-ffuf-res}
\end{figure*}

Figure \ref{fig:armageddon-ffuf-res} shows the result of executing \texttt{ffuf} in port 80 of Armageddon. The result shows a list of paths that are accessible and this information will help the user to decide if it is necessary to execute \texttt{ffuf} in some new path, or if it is time to use an external tool to continue the exploration of the service.

\section{Conclusion}
\label{sec:conclusion}

In this paper we discussed Shadow Blade: a tool to interact with attack graphs \cite{shdw}. This tool can help cyber security professionals and CTF competitors to execute reconnaissance tools, gather the output data and generate an attack graph in order to show the possible attack vectors that the user can explore. We focused on the presentation of the tool showing that the tool was developed to be used in a real-world scenario and we beta tested it in a CTF example, which is a small and custom computer network.

We also discussed related work of tools that automated all the process of reconnaissance, vulnerability discovery and exploitation of real life computer networks, which could be developed inside Shadow Blade using its modular components.

Shadow Blade was not developed to automate all the reconnaissance, vulnerability analysis and exploitation process, so the cyber security professionals and CTF competitors can sharpen their offensive skills. Therefore, Shadow Blade can perform some automated tasks for reconnaissance to increase the speed on common tasks.

On the other hand, we developed Shadow Blade to be modular enough that cyber security professionals and CTF competitors can develop new functionalities that have a better fit for their workflow. Hence, the general idea is that the user can use Shadow Blade to execute common reconnaissance tasks for an initial analysis and afterwards execute more specialized tools outside, creating new nodes in Shadow Blade to represent the result of the specialized tools.

\section{Future Work}
\label{sec:future}

We expect to expand the \textbf{Weight} module to receive a graph and execute the computation of the weights inside a planner and, probably, add some type of machine learning so that Shadow Blade can learn how to better compute the weight of the edges and propose a better attack graph for the user. Another improvement would be to add more tools in \textbf{Code Runner} so that the user can execute as much tools as it is possible via Shadow Blade, so it can generate a more precise attack graph for the user.

Furthermore, in Section \ref{sec:background} we presented an initial formalization that could be used to better describe our work in a formal way, similar to the work of Peralta \textit{et al.} \cite{Peralta08}.

\printbibliography

\end{document}